\documentclass[prb,showpacs,preprintnumbers,amsmath,amssymb,superscriptaddress]{revtex4}
\usepackage{epsf}
\usepackage{graphicx}
\usepackage{bm}
\usepackage{color}

\begin{document}

\title{Configuration interaction in delta-doped heterostructures}

\author{I.~V.~Rozhansky}
\email{rozhansky@gmail.com} \affiliation{A.F.~Ioffe Physical
Technical Institute, Russian Academy of Sciences, 194021
St.Petersburg, Russia} \affiliation{Lappeenranta University of
Technology, P.O. Box 20, FI-53851, Lappeenranta, Finland}
\author{N.~S.~Averkiev}
\affiliation{A.F.~Ioffe Physical Technical Institute, Russian
Academy of Sciences, 194021 St.Petersburg, Russia}
\author{E.~L\"ahderanta}
\affiliation{Lappeenranta University of Technology, P.O. Box 20,
FI-53851, Lappeenranta, Finland}


\date{\today}

\begin{abstract}
We analyze the tunnel coupling between an impurity state located in
a $\delta$-layer and the 2D delocalized states in the quantum well
(QW) located at a few nanometers from the $\delta$ -- layer. The
problem is formulated in terms of Anderson-Fano model as
configuration interaction between the carrier bound state at the
impurity and the continuum of delocalized states in the QW. An
effect of this interaction on the interband optical transitions in
the QW is analyzed. The results are discussed regarding the series
of experiments on the GaAs structures with a $\delta$-Mn layer.
\end{abstract}

\maketitle

\section{Introduction}
\label{secIntro} The problem of so-called configuration interaction
of a single bound state with a continuum of states goes back to the
famous paper by U. Fano \cite{PhysRev.124.1866} rated as one of the
most relevant works of 20th century.\cite{RevModPhys.82.2257} The
suggested theoretical approach often regarded as Fano-Anderson model
or configuration interaction succeeded in explaining puzzling
asymmetric resonances observed in various experiments in atomic and
nuclear physics, condensed matter physics and optics.
\cite{RevModPhys.82.2257}. The co-existence of the discrete energy
level and the continuum states within the same energy range is also
quite common in low-dimensional semiconductor structures.
\cite{RevModPhys.82.2257,PhysRevB.65.155302,okulov:220,springerlink:10.1134/S1063782608080034}
Of particular interest nowadays are the structures having a quantum
well (QW) and a ferromagnetic or paramagnetic layer located in the
vicinity of the QW, but not penetrating into the QW region. In such
structures high mobility of the carriers along the QW is combined
with the magnetic properties provided by the magnetic layer. A
number of recent experiments show that the Mn $\delta$--layer gives
rise to circular polarization of the photoluminescence (PL) from the
QW in an external magnetic field applied perpendicular to the QW
plane.\cite{springerlink:10.1134/S1063783410110144,springerlink:10.1134/S0021364009220056}
It was questioned whether the spin polarization of the carries in
the QW is due to the electrons tunneling to Mn site or the tunnel
coupling of the holes at Mn with those in the QW. The latter
mechanism seemed to lack the proper theoretical description. In this
paper we try to fill this gap. We show that the simple scheme of the
holes configuration interaction leads to the opposite sign of the
circular polarization than that observed in the experiment. The
model system considered in the present paper consists of a
$\delta$--layer of the impurities (donors or acceptors) and a QW
having one level of size quantization for the electrons or holes
respectively. The energy level of the impurity bound state lies
within the range of the 2D states size quantization subband in the
QW. We will be considering the case of rather deep impurity level in
the sense that the impurity activation energy substantially exceeds
the kinetic energy of the 2D carriers in the QW. The attracting
potential of the impurity is assumed spherically symmetric and since
it is a deep level we treat it with zero radius potential
approximation\cite{Lucovsky}. At that we consider both the simple
band structure and the one of the GaAs valence band type.

\section{Tunneling between impurity and quantum well}
\label{secT} In this section we consider the configuration
interaction between a single impurity bound state and the continuum
of 2D states in the QW. The potential barrier separating the
impurity from the QW is assumed to be weakly transparent for the
tunneling. Rigorous calculation of the eigenfunctions is rather hard
to perform as it requires solving stationary Schrodinger equation in
the complicated 3D potential. In order to circumvent the explicit
solving of the Schrodinger equation for tunneling problems the
so-called tunneling or transfer Hamiltonian formalism is commonly
used as originally proposed by Bardeen \cite{PhysRevLett.6.57}. The
total Hamiltonian is expressed as $H=H_{i}+H_{QW}+H_T$, where
$H_{i}$ is partial Hamiltonian having the bound state at the
impurity as its eigen state. $H_{QW}$ in the same way corresponds to
the QW itself, its eigenfunctions $\varphi_{\lambda}$ form
non-degenerate continuum of states characterized by the quantum
number(s) $\lambda$. The term $H_T$ accounts for the tunneling. In
the secondary quantization representation the total Hamiltonian can
be written as follows:
\begin{equation}
\label{eqTunHamil} H = {\varepsilon_0}{a^ + }a + \int {{\varepsilon
_{\lambda}}c_{\lambda}^ + {c_{\lambda}}d\lambda + } \int
{\left({t_{\lambda}}c_{\lambda}^ + a + t_{\lambda}^*{a^ +
}{c_{\lambda}}\right)d\lambda},
\end{equation} where $a^+, a$ --
the creation and annihilation operators for the bound state
characterized by its energy $\varepsilon_0$, and $c^+_{\lambda},
c_{\lambda}$ -- the creation and annihilation operators for a
continuum state having energy $\varepsilon _{\lambda}$. The energy
here and below is measured from the level of size quantization of
the carriers in the QW so that $\varepsilon _{\lambda}$ is simply
their kinetic energy.
 The expression (\ref{eqTunHamil}) is rather general, in fact
it can be regarded as introducing the coupling between two systems
into the Hamiltonian in the most simple phenomenological way. From
this viewpoint the coupling parameter $t_{\lambda}$ is still to be
determined through exact solving of the eigenvalue problem for the
whole system. Bardeen's approach suggests a simple recipe for
calculation of the tunneling parameter for the case of weak
tunneling through a potential barrier:
\begin{equation}
\label{eqTunel} {t_{\lambda}} =
\int_{a}{\left({\varphi_{\lambda}^*}K{\psi} -
{\psi}K{\varphi_{\lambda}^*}\right)}d\mathbf{r}  ,
\end{equation}
where integration is performed over region $a$ to the one side of
the barrier. Here $K$ is the kinetic energy operator:
 \begin{equation}
\label{eqkinEn}
 K =  - \frac{{{\hbar ^2}}}{{2m}}\Delta .\end{equation}
The attraction potential of the impurity is considered spherically
symmetric, so the whole system (impurity+QW) has the cylindrical
symmetry with $z$ axis directed normally to the QW plane and going
through the impurity center. Thus for further calculations it will
be most convenient to represent the QW states in cylindrical
coordinates rather than as plane waves. In this case each state is
characterized by the wavenumber $k$ and the cylindrical harmonic
number $l$:
\begin{equation}
\label{eqwavefuncylindr} \varphi_{kl}=\eta \left( z \right)\sqrt
{\frac{{m}}{{2\pi {\hbar ^2}}}} {J_l}\left( {k\rho } \right)
e^{il\theta}
      \end{equation}
where $J_l(k\rho)$ is the Bessel function of order $l$, $\rho$ and
$\theta$ are the polar coordinates in the QW plane,  $m$--the
in-plane effective mass, $\eta \left( z \right)$ is the envelope
function of size quantization in $z$-direction.
 The wavefunction (\ref{eqwavefuncylindr}) has the normalization:
\begin{equation} \label{eqNormDelta}
 \left\langle {\varphi_{kl}|\varphi_{k'l'}} \right\rangle =
\delta\left(\varepsilon  - \varepsilon '\right)\delta_{ll'},
\end{equation}
where $\varepsilon=\hbar^2 k^2/2m$. The potential barrier separating
the deep impurity level from the QW in the first approximation can
be assumed having a rectangular shape. Inside the barrier the
function $\eta(z)$ is (z-axis is directed towards the impurity,
$z=0$ corresponds to the QW boundary):
\begin{equation}
\eta \left( z \right)\sim \frac{1}{\sqrt{a}}e^{-qz},
\end{equation}
where $q=\sqrt{\frac{{2m{E_0}}}{{{\hbar ^2}}}}$, $a$ is the QW
width, $E_0$ is the binding energy of the bound state, at the same
time $E_0$ determines the height of the potential barrier. Let us
firstly consider the simple band case valid for the bound electrons
at donor impurity coupled to the QW conductance band. The spherical
potential of the impurity results in the ground state of the carrier
to be angular independent, therefore the efficient tunneling overlap
occurs only with the zeroth cylindrical harmonic
$\varphi_{k0}\equiv\varphi(\varepsilon)$. For the deep impurity
level one can use zero radius potential approximation\cite{Lucovsky}
and express the s-type wavefunction as:
\begin{equation}
\psi={\sqrt{2q}}\frac{e^{-qr}}{r}.
\end{equation}
The integration (\ref{eqTunel}) over the space is reduced to the
integration over the surface $\Omega_S$ inside the barrier which is
more convenient to take at the impurity site. This yields for the
electrons tunneling between the donor state and the QW:
\begin{equation}
\label{eqTe} t_k^{e}  =
\sqrt{\frac{2\pi}{{aq}\left(1+\frac{k^2}{q^2}\right)}}\sqrt{E_0}e^{-qd}
\end{equation}
It is clearly seen that as long as the case $k<<q$ is considered,
the tunneling parameter has very weak dependence on $k$.
 In order to apply the same approach to the holes
tunneling in GaAs it has to be generalized for the case of the
valence band complex structure. Let us consider In$_{x}$Ga$_{1-x}$As
QW having only one level of size quantization for the heavy holes
and neglect the light holes being split off due to the size
quantization. The basis of Bloch amplitudes to be used is formed of
the states with certain projection of the total angular momentum
$J=3/2$ on $z$ axis.
 It would be
tempting to generalize (\ref{eqTunel}) by treating $K$ as the
kinetic part of the Luttinger Hamiltonian ($\hbar k_x$,$\hbar
k_y$,$\hbar k_z$ are, as usual, the momentum operators along the
appropriate axis):
\begin{equation} \label{eqLuttinger} K = \left(
{\begin{array}{*{20}{c}}
   F & H & I & 0  \\
   {{H^*}} & G & 0 & I  \\
   {{I^*}} & 0 & G & { - H}  \\
   0 & {{I^*}} & { - {H^*}} & F  \\
\end{array}} \right),
\end{equation}

\begin{align}
 F &=  - A{k^2} - \frac{B}{2}\left( {{k^2} - 3k_z^2} \right), \nonumber \\
 G &=  - A{k^2} + \frac{B}{2}\left( {{k^2} - 3k_z^2} \right), \nonumber \\
 H &= D{k_z}\left( {{k_x} - i{k_y}} \right) \nonumber, \\
 I &= \frac{{\sqrt 3 }}{2}B\left( {k_x^2 - k_y^2} \right) -
 iD{k_x}{k_y}.
 \end{align}
The functions $\psi_{\alpha}$, $\varphi_{\lambda \beta}$ in
(\ref{eqTunel}) become now 4-component vector functions (also the
spin indices $\alpha$ and $\beta$ are added here). The explicit
expression for the bound hole state functions $\psi_{\alpha}$ and
the 2D hole states $\varphi_{\lambda\beta}$ can be found in
Ref.\cite{PhysRevB.85.075315}. The important thing about those is
while the decay length in $z$--direction of the 2D wavefunctions
$\varphi_{\lambda\beta}$ is controlled by the heavy hole mass
$m_{hh}\approx0.5\;m_0$ ($m_0$ is the free electron mass), the decay
length of radial part of the bound state wavefunction
$\psi_{\alpha}$ is characterized by both heavy hole mass $m_{hh}$
and the light hole mass $m_{lh}\approx0.08\;m_0$
\cite{PhysRevB.85.075315}. Analogously to the simple band case the
integration (\ref{eqTunel}) over the whole space is reduced to the
integration over the surface $\Omega_S$ inside the barrier, at that,
only $z$--projection of the kinetic energy operator is required. The
expression for tunneling parameter simplifies into:
\begin{equation}
\label{eqTLuttinger} {t_{kl\alpha\beta}}^{(h)} = \left( {B - A}
\right)\int_{\Omega_S} {dS} \left( {{\varphi
_{kl\beta}}^*\frac{d}{{dz}}{\psi_{\alpha}} - {\psi_{\alpha}
}\frac{d}{{dz}}{\varphi _{kl\beta}}^*} \right),
\end{equation}
where $\varphi _{kl}$ is given by (\ref{eqwavefuncylindr}).

Regrettably, the above given straightforward generalization of
(\ref{eqTunel}) fails to be fully correct. Indeed, the largest decay
length of the bound state $\psi_{\alpha}$ is determined by the light
hole mass while the decay length of the QW states is governed by the
heavy hole. Due to this circumstance the result of the surface
integration (\ref{eqTLuttinger}) becomes dependent on the particular
position of the integration surface inside the barrier. However, it
can be shown that in the case of two masses the exponential
dependence of the tunneling parameter on the barrier thickness is
determined by the smallest mass, but the exact value of the
tunneling parameter cannot be correctly obtained within the given
approach. Now we define $q=\frac{\sqrt{2m_{hh}E_0}}{\hbar^2}$,
$\beta=m_{lh}/m_{hh}$. The explicit evaluation of the overlap
integrals with account for $k<<q$ shows that the tunneling
configuration interaction to be accounted for is only between the
zeroth cylindrical harmonic $\varphi_{k0,-\frac{3}{2}}$ and the
bound state $\psi_{-\frac{3}{2}}$ as well as between
$\varphi_{k0,+\frac{3}{2}}$ and $\psi_{+\frac{3}{2}}$. Both are
governed by the same tunneling parameter $t_k^h$:
\begin{align}
\label{eqT}
 & t_k^{h}  =  \left( \frac{A - B}{\hbar^2/2m_0} \right)\sqrt{\frac{\pi}{aq}}\sqrt{\frac{{ m_{hh}m'_{hh}}}{{m_0^2} }}
\zeta  \left( {k/q} \right)\beta\sqrt{E_0}\exp{\left( -\chi \left(
{k/q} \right)\sqrt{\beta} qd \right)},
\end{align}
where $1\leq \chi\leq2$, $\zeta\sim 1$ are weak dimensionless
functions of $k/q$, $m'_{hh}$ is the effective in-plane heavy hole
mass. The tunneling parameter $t_k^h$ exponentially depends on the
barrier thickness with the light hole mass entering the exponent
index. The particular expressions for $\chi$ and $\zeta$ depend on
the surface one chooses for the integration in (\ref{eqTunel}).

In both cases for $t_k^{e},t_k^{h}$ it is reasonable to assume that
the tunneling parameter does not depend on $k$ as weak tunneling
implies $k<<q$. Still, its rapidly decreasing behavior for $k>>q$
has to be kept in mind when it provides convergence for integration
over $k$. In our estimations the shape of the potential barrier
separating the QW was assumed rectangular. This is quite reasonable
for the estimation at $k<<q$. However, the particular shape of the
barrier becomes important when one is concerned with experimental
dependence on the distance $d$ between the impurity and the QW.

\section{Effect on the luminescence spectrum}
The transfer Hamiltonian (\ref{eqTunHamil}) with known tunneling
parameter $t(\varepsilon)$ allows one to construct the
eigenfunctions $\Psi$ of the whole system given those of the bound
state $\psi$ and the QW states $\varphi(\varepsilon)$ :
\begin{equation} \label{eqExpand} \Psi \left( E \right) = {\nu
_0}\left( E \right){\psi} + \int_0^\infty {\nu \left( {E,\varepsilon
} \right)} \varphi \left( \varepsilon \right)d\varepsilon,
\end{equation}
$E$ denotes the energy of the state $\Psi$. Here
$\varphi(\varepsilon)$ are the wavefunctions with zeroth cylindrical
harmonic, as was shown above the other harmonics are not affected by
the tunneling configuration interaction.
 Plugging
(\ref{eqExpand}) into the stationary Schrodinger
 equation:
\[H\Psi=E\Psi
\] with $H$ being the effective Hamiltonian (\ref{eqTunHamil}) one
gets the following system of equations:
 \begin{equation}
 \label{eqFanoSystemContinous}
 \begin{array}{l}
 {\nu _0}\left(E\right){\varepsilon _0} + \int_0^\infty {t\left( \varepsilon  \right)\nu \left( E,\varepsilon  \right)d\varepsilon }  = E{\nu _0}\left(E\right),\,\,\, \\
 \nu \left( E,\varepsilon  \right)\varepsilon  + t\left( \varepsilon  \right){\nu _0}\left(E\right) = E\nu \left( E,\varepsilon  \right).\,\,\,\,\,\,\,
 \end{array}
 \end{equation}
In the present work we consider the case of the bound level energy
lying within the range of the continuum: $\varepsilon_0>>t^2$. For
this case the solution is obtained as shown in
\cite{PhysRev.124.1866}:
\begin{align}
\label{eqSolution} & {\nu _0}^2\left( E \right) = \frac{{{t^2}\left(
E \right)}}{{{\pi ^2}{t^4}\left( E \right) + {{\left( {E -
\widetilde{{\varepsilon
_0}}} \right)}^2}}}, \nonumber \\
&\nu \left( E,\varepsilon \right) = {\nu _0}\left( E \right)\left(
{P\frac{{t\left( \varepsilon \right)}}{{E - \varepsilon }} +
Z\left(E\right)t\left(E\right)} \delta \left( {E - \varepsilon }
\right)\right),
\end{align}
 where
\begin{align}
\label{eqZ} Z\left( E \right) &= \frac{E - \varepsilon_0 - F\left(E
\right)}{t^2\left(E \right)}, \nonumber \\
F\left( E \right) &= \int_0^\infty {P\frac{{{t^2}\left( \varepsilon
\right)}}{{\left( {E - \varepsilon } \right)}}d\varepsilon},
\end{align}
 $P$ stands for the principal value and $\tilde \varepsilon_0$ is the center of configuration
 resonance, which appears to be slightly shifted from
 $\varepsilon_0$:
\begin{equation}
\label{eqResonance} \tilde \varepsilon_0(E)=\varepsilon _0 + F(E).
\end{equation}
Because of $k<<q$ it is reasonable to put $t=$const everywhere,
except for (\ref{eqZ}) where decrease of $t$ at $E\rightarrow\infty$
is necessary for convergence of the integral.

In order to analyze the influence of the configuration interaction
on the luminescence spectra we have to calculate matrix element of
operator $\hat{M}$ describing interband radiative transitions
between the hybridized wavefunction $\Psi(E)$ and wavefunction of 2D
the carrier in the other band of the QW which we denote by
$\xi_{k'l'}$, here $k'$ is the magnitude of the wavevector, $l'$ is
the number of cylindrical harmonic analogously to
($\ref{eqwavefuncylindr}$). If, for instance, one considers the
acceptor-type impurity then $\Psi(E)$ is the hybridized wavefunction
of the 2D holes and $\xi_{k'l'}$ is the wavefunction of the 2D
electrons in the QW.  We assume that (a) there are no radiative
transitions between the bound state wavefunction $\psi$ and the 2D
carrier wavefunction $\xi_{k'l'}$ in the other band thus the matrix
element for transitions from the bound state:
\begin{equation}
\label{eqM0} \left\langle {\xi_{k'l'}\left| {\hat M} \right|\psi}
\right\rangle =0,\end{equation} (b) the interband radiative
transitions between the free 2D states in the QW are direct.
According to (\ref{eqwavefuncylindr}) the wavefunctions
$\varphi(\varepsilon)$ and $\xi(\varepsilon')$ corresponding to the
zeroth harmonic in the cylindrical basis are: \begin{align}
\varphi(\varepsilon)=\eta(z)\sqrt {\frac{{{m}}}{{2\pi {\hbar ^2}}}}
{J_0}\left( {{k}\rho } \right) \nonumber\\
\xi(\varepsilon')=\zeta(z)\sqrt {\frac{{{m'}}}{{2\pi {\hbar ^2}}}}
{J_0}\left( {{k'}\rho } \right),\end{align} where
\[{k} = \frac{{\sqrt {2{m}\varepsilon } }}{\hbar },\,\,\,{k'} = \frac{{\sqrt {2{m'}\varepsilon' } }}{\hbar },\]
$\eta(z),\zeta(z)$ -- the appropriate size quantization functions in
z--direction, $m$, $m'$ are the in-plane masses of the electrons and
holes respectively if the donor-type impurity is considered and vice
versa for the acceptor case.
 Without the tunnel coupling the matrix
element for the direct optical transitions between the states
$\varphi(\varepsilon)$ and $\xi(\varepsilon')$ is given by:
\begin{equation} \label{eqMatrElement}
M_{0}\left(\varepsilon,\varepsilon'\right)=  \left\langle
{\xi(\varepsilon')\left| {\hat M} \right|\varphi({\varepsilon})}
\right\rangle =u_k\frac{{\sqrt {m m'} }}{{k{\hbar ^2}}}\delta \left(
{k - {k'}} \right) ,
\end{equation}
where $u_k$ is the appropriate dipole matrix element for the Bloch
amplitudes. According to the above mentioned considerations it is
only this matrix element that is affected by the tunnel coupling,
preserving the matrix elements corresponding to the transitions
between other than the zeroth cylindrical harmonic. Denoting by $M$
the modified matrix element for transitions between the states
$\Psi(E)$ and $\xi(\varepsilon')$ with the further use of the Fano
theory\cite{PhysRev.124.1866} one obtains:
\begin{equation}
\label{eqM} {{M\left(E,\varepsilon'\right)^2
}}={{M_0\left(E,\varepsilon'\right)^2 }} \left[ 1 - \frac{{\pi ^2
t^4 }}{{\pi ^2 t^4 + \left( {E - \widetilde{\varepsilon _0 }}
\right)^2 }}\right]
\end{equation}
We proceed further with the Fermi's Golden Rule for the transition
probability:
\begin{equation} \label{eqFermi} W(\hbar \omega ) = \frac{{2\pi
}}{\hbar }\int_0^\infty\int_0^\infty {\left|M \left( {E,\varepsilon'
 }\right)\right|^2 } f' \left( {\varepsilon' } \right)f \left( E
\right)\delta \left( {E + \varepsilon'  + E_g - \hbar \omega }
\right)dE d\varepsilon',
\end{equation} where $E_g$ -- the QW bandgap,
$\hbar\omega$ -- the energy of the radiated photon, $f, f'$ -- the
energy distribution functions for the carriers in the hybridized and
intact bands respectively. Substituting (\ref{eqMatrElement}) and
(\ref{eqM}) into (\ref{eqFermi}) one should treat correctly the
delta-function for the wavenumbers of the zeroth cylindrical
harmonic. It can be shown that:
\[
 \delta^2(k-k')=\frac{\sqrt{S}}{\pi^{3/2}}\delta(k-k'),
\]
where $S$ is the area of the QW. Then we arrive at:
\begin{equation}
\label{eqW} W\left( {\hbar \omega } \right) = \frac{{{u^2}f\left(
E_\omega \right)}}{{{\pi ^{1/2}}{\hbar ^2}}}\frac{{\sqrt
{2\widetilde{m}S} }}{{\sqrt {\hbar \omega  - {E_g}} }}\left( {1 -
\frac{{{\pi ^2}{t^4}}}{{{\pi ^2}{t^4} + {{\left( {{E_\omega } -
\widetilde{{\varepsilon _0}}} \right)}^2}}}} \right),\end{equation}
where
\begin{align}
\label{eqEomega} & f\left( {{E_\omega }} \right) = {f'  }\left(
{\alpha^{-1} {E_\omega }}
\right){f}\left( {{E_\omega }} \right), \nonumber \\
&E_\omega = \frac{\hbar\omega - E_g}{1 + \alpha ^{-1}},
\nonumber \\
& \widetilde{m}=\frac{mm'}{m+m'},\nonumber \\
& \alpha=m'/m,\nonumber \\
 \end{align}
while for the all cylindrical harmonics altogether the unperturbed
optical transition rate yields: \begin{equation}
\label{eqW0}{W_{0}}\left( {\hbar \omega } \right) = \frac{{2\pi
{u^2}f\left( E_\omega \right)}}{\hbar }\left(
{\frac{{\widetilde{m}}}{{{\hbar ^2}}}S} \right).\end{equation}
 The result (\ref{eqW}) obtained for a single impurity can be applied to
an ensemble of impurities provided their interaction between each
other is weak compared to the tunnel coupling with the QW. In this
case the sample area $S$ should be replaced with $n^{-1}$, $n$ being
the sheet concentration of the impurities in the delta-layer. After
normalization by the area of the QW from (\ref{eqW}),(\ref{eqW0}) we
finally get the spectral density of the luminescence intensity:
\begin{equation}
\label{eqInt} I\left( {\hbar \omega },\widetilde{{\varepsilon _0}}
\right) = {I_0}\left( {\hbar \omega } \right)\left( {1 -
a\left(\widetilde{{\varepsilon _0}}\right)\sqrt{n}\frac{{{\pi
^2}{t^4}}}{{{\pi ^2}{t^4} + {{\left( {{E_\omega } -
\widetilde{{\varepsilon _0}}} \right)}^2}}}} \right),
\end{equation} where
\[a(\widetilde{{\varepsilon _0}}) = \frac{\hbar }{{{\pi ^{3/2}}\sqrt {2\widetilde{m}\widetilde{{\varepsilon _0}}\left( {1 + {\alpha ^{ - 1}}} \right)} }},\]
\[{I_0}\left( {\hbar \omega } \right) = \frac{{2\pi {u^2}\widetilde{m}}}{{{\hbar ^3}}}f\left( {{E_\omega }} \right).\]

\section{Polarization of the spectra}
\label{SecPolar} It follows from (\ref{eqInt}) that the bound state
lying within the energy range of the continuum causes a dip in the
luminescence spectra emitted from the QW. If then for any reason the
bound state is split the luminescence spectra will show the
appropriate number of the dips shifted by the splitting energy
$\Delta$. If one considers the splitting in the magnetic field
applied along $z$ each of the split sublevels is characterized by
certain projection of spin and interacts with only one of the 2D
carriers spin subbands characterized by the same projection of spin.
Thus, for each of the two circular polarizations $\sigma^+$,
$\sigma^-$ of the light emitted from the QW one would expect one
dip, its spectral position being different for $\sigma^+$ and
$\sigma^-$ in accordance with the splitting energy $\Delta$. As an
example let us consider the GaAs-based QW and 2D heavy holes
interacting via the tunneling configuration interaction with the
bound state at an acceptor. This case is shown schematically in
Fig.1. The 2D holes with the projections of total angular momentum
$j=+3/2$ and $j=-3/2$ recombine emitting respectively right-
($\sigma^+$) and left- ($\sigma^-$) circularly polarized light. In
section \ref{secT} it was shown that the heavy holes with
$j=-3/2(j=+3/2)$ interact basically with the bound states
$\psi_{-\frac{3}{2}}(\psi_{+\frac{3}{2}})$. An external magnetic
field applied along $z$ would cause Zeeman splitting of the bound
state energy level $\varepsilon_0$ into
$\varepsilon_0^+=\varepsilon_0+\Delta/2$ and
$\varepsilon_0^-=\varepsilon_0-\Delta/2$.
 The splitting
$\Delta=\varepsilon_0^+-\varepsilon_0^-$ may also originate from
exchange interaction of the holes with spin-polarized acceptor ions.
Let us refer to the case of Mn ions having positive g-factor
($g\approx3$, see Ref.\cite{PhysRevLett.59.240}). The hole is
coupled to Mn in antiferromagnetic way thus the level
$\varepsilon_0^+$ corresponds to $j=-3/2$ and $\varepsilon_0^-$ to
$j=+3/2$.  As follows from (\ref{eqResonance}),(\ref{eqEomega}) the
difference in the positions of the resonances (dips) $E^+_\omega$
and $E^-_\omega$ corresponding to the bound state sublevels
$\varepsilon_0^+$ and $\varepsilon_0^-$ is given by:
\begin{equation}\label{eqdeltatilde}\widetilde\Delta  = {E^ +_\omega } - {E^ -_\omega } = \Delta  + {t^2}\ln
\left( {1 + \frac{{\widetilde\Delta }}{{{E^-_\omega }}}}
\right).\end{equation} Unless the positions of the resonances are
too close to the band edge the last term in (\ref{eqdeltatilde}) can
be neglected and $\widetilde{\Delta}=\Delta=
    \varepsilon_0^+-\varepsilon_0^-$. With account for the energy
distribution functions for the holes and electrons the shifted
positions of the resonances lead to the difference in the
luminescence intensity for the opposite circular polarizations. In
the discussed example of the antiferromagnetic alignment of the hole
the luminescence spectra $I^+(\hbar\omega,\widetilde{{\varepsilon
_0}}^+)$, $I^-(\hbar\omega,\widetilde{{\varepsilon _0}}^-)$ having
the resonance positions at $\varepsilon _0^+$ and $\varepsilon _0^-$
correspond to the circular polarizations $\sigma^-$  and $\sigma^+$
respectively. As can be seen from (\ref{eqInt}) the difference in
the resonance positions $\Delta=\widetilde{{\varepsilon
_0}}^+-\widetilde{{\varepsilon _0}}^-$ leads to the integral
polarization of the spectra if the distribution function $f(E)$
significantly varies in the vicinity of $\varepsilon _0$. This is
illustrated in Fig.2. The functions $I^-$ and $I^+$  are shown by
blue and red solid lines respectively. The integral polarization is
naturally defined as:
\[P = \frac{P(\sigma^+)-P(\sigma^-)}{P(\sigma^+)+P(\sigma^-)}\approx\frac{{\int\limits_{{E_g}}^\infty  {{I^- }\left( {\hbar \omega } \right)d\left( {\hbar \omega } \right)}  - \int\limits_{{E_g}}^\infty  {{I^+ }\left( {\hbar \omega } \right)d\left( {\hbar \omega } \right)} }}{{2\int\limits_{{E_g}}^\infty  {{I_0}\left( {\hbar \omega } \right)d\left( {\hbar \omega } \right)} }}\]
With use of (\ref{eqInt}) this yields: \begin{equation}
\label{eqPolInt} P = -\sqrt{n}\frac{{ \int\limits_0^\infty\pi
t^2\left(E\right) {\left[ {\frac{a\left(\widetilde{\varepsilon} _0^
+ \right){{\pi t^2}\left( E \right)}}{{{\pi ^2}{t^4}\left( E \right)
+ {{\left( {E - {\rm{ }}\widetilde{\varepsilon} _0^ + }
\right)}^2}}} - \frac{a\left(\widetilde{\varepsilon} _0^ -
\right){{\pi t^2}\left( E \right)}}{{{\pi ^2}{t^4}\left( E \right) +
{{\left( {E - {\rm{ }}\widetilde{\varepsilon} _0^ - } \right)}^2}}}}
\right]f\left( E \right)dE} }}{{2\int\limits_0^\infty {f\left( E
\right)dE} }}.\end{equation}
 The slow varying functions $f(E)$ and $\widetilde{\varepsilon}_0(E)$ in the integral may be assumed
as constants taken at $\widetilde{\varepsilon}_0^-,
\widetilde{\varepsilon}_0^+$, the tunneling parameter will be
treated as a constant in the whole range of interest $t^2(E)\equiv
t^2$.

Then treating the expression in brackets as delta-functions we
obtain:
\begin{equation}
\label{eqPolGeneral} P = -\frac{\sqrt{\pi}{\hbar {t^2}\sqrt n }}{{{{
{2} }^{3/2}}\sqrt {m} }}\frac{{f\left( {\varepsilon _0^ + }
\right){{\left( {\varepsilon _0^ + } \right)}^{ - 1/2}} - f\left(
{\varepsilon _0^ - } \right){{\left( {\varepsilon _0^ - } \right)}^{
- 1/2}}}}{{\int\limits_0^\infty  {f\left( E \right)dE} }}.
\end{equation}
Note that for the considered example the polarization degree appears
to be negative. The positive sign would have appeared if the
ferromagnetic coupling between the acceptor ion and the hole had
been assumed.

\section{The electrostatic effect}
\begin{figure}
  \leavevmode
 \centering\includegraphics[width=0.4\textwidth]{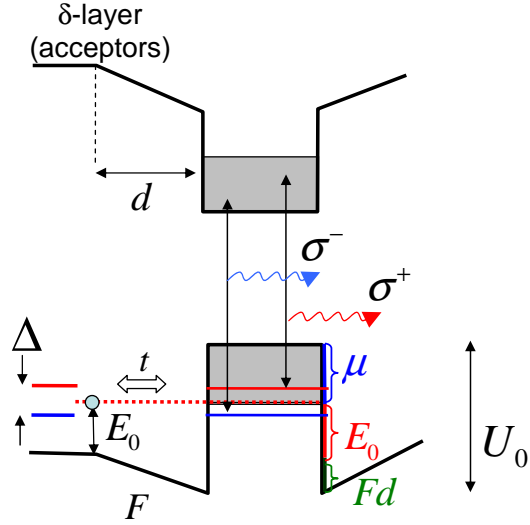}
 \caption{(Color online) Mechanism of polarization of the luminescence for the acceptor type impurity. The localized hole levels split
 in magnetic field. Each of them effectively couples with
 the 2D holes having certain projection of angular momentum. Shifted positions of the resonances
 with account for temperature distribution of the holes cause the difference
 in intensities of circular polarizations $\sigma^+$, $\sigma^-$. The scheme also shows the simple electrostatic model described in the text.}
 \label{fig_scheme}
\end{figure}

Because of the tunneling involved in the polarization of the
luminescence one might reasonably expect very strong dependence of
the polarization degree on the distance $d$ between the
$\delta$--layer and the QW (i.e. the thickness of the spacer).
However, the purely exponential dependence of the polarization on
the barrier thickness appears to be weakened due to the
electrostatic effect shown in Fig.\ref{fig_scheme} and explained
below. Let us for simplicity consider the electrons distribution
function being nearly constant within the configuration resonances.
The holes are considered to have Fermi distribution function
characterized by the chemical potential $\mu$ and the temperature
$T$. In the absence of an external optical pumping the holes in the
QW are in thermodynamic equilibrium with the acceptors in the
$\delta$--layer, therefore they have the same chemical potential.
Under low pumping conditions the already large concentration of the
holes in the QW is not strongly violated, so it is reasonably to
assume that the quasi Fermi levels of the holes at the acceptors and
in the QW coincide, it means that $\varepsilon_0=\mu$. Strictly
speaking, this is valid for a single bound level, if the level is
split so that $\varepsilon_0^+-\varepsilon_0^-= \Delta$, one should
probably assume $\varepsilon_0^-=\mu$. From (\ref{eqPolGeneral}) we
get the following simplified expression:
\begin{equation}
\label{eqPoltanh} P = -\frac{{\sqrt{\pi}\hbar {t^2}\sqrt n
}}{{{2^{5/2}}{}\sqrt {m'_{hh}} {\mu ^{3/2}}}}\tanh \frac{\Delta
}{{2kT}}
\end{equation}
As we will show below both $t$ and $\mu$ contribute to the
dependence of the integral polarization $P$ on the spacer thickness
$d$ and the QW depth $U_0$.
 The holes in the QW provide an
electrical charge density estimated as $\sigma=eN\mu$, where $e$ is
the elementary charge, $N$ is the 2D density of states. The
positively charged plane of the QW and negatively charged
$\delta$--layer of partly ionized acceptors separated by a distance
$d$ produce an electric field
\begin{equation}\label{eqF}F=\frac{4\pi
eN\mu}{\varepsilon},\end{equation} $\varepsilon$ being dielectric
constant of the material. Due to the electric field $F$ the valence
band edge at position of Mn layer appears to be shifted from the
valence band edge just outside of the QW by $F\cdot d$. Because the
quasi Fermi level of the acceptors exceeds the local position of the
valence band edge by the binding energy $E_0$, the equality of the
quasi Fermi levels leads to a simple equation (see
Fig.\ref{fig_scheme}) :
\begin{equation} U_0 = \mu + E_0 + eFd,
\end{equation}
where $U_0$ is the QW depth and $\mu$ is the chemical potential
 of the holes in the QW. With (\ref{eqF}) one gets :
\begin{equation}
\label{eqmudependence} \mu  = \frac{{U_0  - E_0 }}{{1 + \frac{{4\pi
Ned}}{\varepsilon }}} \approx \frac{{\left( {U_0  - E_0 }
\right)\varepsilon }}{{4\pi Ned}}.
\end{equation}
In order to estimate the dependence of the tunneling parameter $t$
on the QW and spacer parameters we consider the WKB tunneling
through trapezoid barrier as seen in Fig.\ref{fig_scheme}. With
taking into account (\ref{eqT}) and (\ref{eqmudependence}) this
leads to the following expression (we assume $\mu\ll U_0$):
\begin{equation}
\label{eqtd} t^2\sim\exp\left(-\kappa d\right),
\end{equation}
where
\begin{equation}
\label{eqkappa} \kappa=\frac{4\sqrt{2m_{lh}}}{3
\hbar\left(U_0-E_0\right)}\left(U_0^{3/2}-E_0^{3/2}\right)
\end{equation}
From (\ref{eqPoltanh}), (\ref{eqtd}), (\ref{eqkappa}),
(\ref{eqmudependence}) follows the dependence of integral
polarization on the spacer thickness: \begin{equation} \label{eqPd}
P\sim d^{3/2}\exp(-\kappa d),
\end{equation}
Note that electrostatic effect results in the dependence of $\mu$ on
$d$ which leads to the dependence of $P$ on $d$ being not purely
exponential but weakened by the pre-exponential factor $d^{3/2}$.
While the correction is pre-exponential it appears to be significant
enough up to  $\kappa d\approx2-3$ which is typical for the
experimental situation.

\section{Discussion}
\label{SecDis} In the proposed theory the polarization of light
emitted from the QW originates from the splitting of the impurity
bound state and therefore may exceed the polarization degree
expected from an intrinsic g-factor of the 2D carriers located in
the QW. 
The sign of the polarization deserves special discussion. As was
shown above, the tunnel coupling causes a dip in the luminescence
spectra. This means that in the considered scheme the polarization
of the luminescence from the QW is expected to be of the opposite
sign than that due to the optical transitions between the bound
state and the free carriers inside the barrier. In particular, the
configuration interaction between the 2D heavy holes and Mn
$\delta$-layer considered in Sec.\ref{SecPolar} leads to the
negative sign of the polarization (a mistake made
in\cite{PhysRevB.85.075315} has mislead to the positive sign). Such
result contradicts the known experimental data
\cite{zaitsev:399},\cite{Korenev}, where the polarization is shown
to be positive. This might suggest that regarding these particular
experiments the polarization is not due to the holes configuration
interaction but rather due to polarization of the electrons as
suggested in \cite{Korenev}. The other possibility might be that the
relevant bound state of the hole at Mn is more complex and does not
resemble the simple antiferromagnetic exchange coupling with Mn ion.

Let us estimate the expected magnitude of the circular polarization
degree due to the tunneling configuration interaction. We assume the
deep impurity level $E_0=100$ meV, the barrier thickness $d=5$ nm,
the QW width $a=10$ nm. Taking  the effective mass as that of the
electrons in GaAs $m=0.06\,m_0$ for the simple band case described
by (\ref{eqTe}) one gets for the tunneling parameter
$\left(t^e\right)^2\approx2$ meV. The estimation for the holes
tunneling parameter appears to be far less, taking $m_{hh}=0.5$
$m_0$, $m_{hh}'=0.15$ $m_0$ from (\ref{eqT}) one gets
$\left(t^{h}\right)^2\sim 0.01$ meV. The polarization degree is to
be estimated using (\ref{eqPolGeneral}). We take $\Delta=1$ meV,
$T_e=T_h=20$ K, the sheet concentration of the impurities
$n=10^{13}$ cm$^{-2}$. Then for the case of the donor impurity
$t=t^e$, $\varepsilon_0=4$ meV, $\mu_h=-1$ meV,
$\mu_e=\varepsilon_0^-$, one gets $|P|\approx 40\%$, for the
acceptor impurity $t=t^h$, $\mu_e=-1$ meV, $\varepsilon_0=2$ meV,
$\mu_h=\varepsilon_0^-$ gives $|P|\approx 0.5\%$. An illustration of
the luminescence spectra for the two circular polarizations is
presented Fig.\ref{figSpectra}. For this we used an intermediate
value for the tunneling parameter $t^2=0.3$ meV ($|P|\approx
0.15\%$) and accounted for inhomogeneous broadening of the spectra
by normal distribution of the bandgap $E_g$ with the dispersion
$\sigma=3$ meV (corresponds to the fluctuation of the QW width by
half a monolayer).

\section{Summary}
The presented theory describes the tunnel coupling between a
continuum of states in the QW and an impurity bound state located
outside of the QW. We utilized the well known Fano approach for
calculation of the matrix elements for the direct interband optical
transitions in the QW. For such transitions the tunnel coupling of
the 2D QW states with the impurity states leads to the drop of the
luminescence spectral density at the frequency corresponding to the
configuration resonance. This modification of the spectra leads to
an integral circular polarization of the light emitted from the QW
provided the bound hole state is split in the projection of the hole
angular momentum. The key advantage of the approach used in the
present study is that the unknown eigenfunctions of the system are
expressed through those of the uncoupled states. Given the expansion
(\ref{eqExpand}) any effects on the localized state can be
translated into effects for the whole coupled system. For this
reason it is capable of describing other effects expected in such
systems like anisotropy of the holes g-factor in the QW induced by
the paramagnetic impurity or the indirect exchange interaction
between the bound states provided by the 2D free carriers in the QW.
\section{acknowledgements}
We thank V. D. Kulakovskii for very fruitful discussions and also
express our thanks to B. A. Aronzon, P. I. Arseev, V. L. Korenev, M.
M. Glazov, V. F. Sapega, S. V. Zaitsev for very useful and helpful
comments. The work has been supported by RFBR (grants no
11-02-00348, 11-02-00146, 12-02-00815,12-02-00141), Russian Ministry
of Education and Science (contract N 14.740.11.0892, contract N
11.G34.31.0001 with SPbSPU and leading scientist G.G. Pavlov), RF
President Grant NSh-5442.2012.2.
\bibliography{MnFano}

\begin{thebibliography}{13}
\expandafter\ifx\csname natexlab\endcsname\relax\def\natexlab#1{#1}\fi
\expandafter\ifx\csname bibnamefont\endcsname\relax
  \def\bibnamefont#1{#1}\fi
\expandafter\ifx\csname bibfnamefont\endcsname\relax
  \def\bibfnamefont#1{#1}\fi
\expandafter\ifx\csname citenamefont\endcsname\relax
  \def\citenamefont#1{#1}\fi
\expandafter\ifx\csname url\endcsname\relax
  \def\url#1{\texttt{#1}}\fi
\expandafter\ifx\csname urlprefix\endcsname\relax\def\urlprefix{URL }\fi
\providecommand{\bibinfo}[2]{#2}
\providecommand{\eprint}[2][]{\url{#2}}

\bibitem[{\citenamefont{Fano}(1961)}]{PhysRev.124.1866}
\bibinfo{author}{\bibfnamefont{U.}~\bibnamefont{Fano}}, \bibinfo{journal}{Phys.
  Rev.} \textbf{\bibinfo{volume}{124}}, \bibinfo{pages}{1866}
  (\bibinfo{year}{1961}).

\bibitem[{\citenamefont{Miroshnichenko
  et~al.}(2010)\citenamefont{Miroshnichenko, Flach, and
  Kivshar}}]{RevModPhys.82.2257}
\bibinfo{author}{\bibfnamefont{A.~E.} \bibnamefont{Miroshnichenko}},
  \bibinfo{author}{\bibfnamefont{S.}~\bibnamefont{Flach}}, \bibnamefont{and}
  \bibinfo{author}{\bibfnamefont{Y.~S.} \bibnamefont{Kivshar}},
  \bibinfo{journal}{Rev. Mod. Phys.} \textbf{\bibinfo{volume}{82}},
  \bibinfo{pages}{2257} (\bibinfo{year}{2010}).

\bibitem[{\citenamefont{Blom et~al.}(2002)\citenamefont{Blom, Odnoblyudov,
  Yassievich, and Chao}}]{PhysRevB.65.155302}
\bibinfo{author}{\bibfnamefont{A.}~\bibnamefont{Blom}},
  \bibinfo{author}{\bibfnamefont{M.~A.} \bibnamefont{Odnoblyudov}},
  \bibinfo{author}{\bibfnamefont{I.~N.} \bibnamefont{Yassievich}},
  \bibnamefont{and} \bibinfo{author}{\bibfnamefont{K.~A.} \bibnamefont{Chao}},
  \bibinfo{journal}{Phys. Rev. B} \textbf{\bibinfo{volume}{65}},
  \bibinfo{pages}{155302} (\bibinfo{year}{2002}).

\bibitem[{\citenamefont{Okulov et~al.}(2011)\citenamefont{Okulov, Lonchakov,
  Govorkova, Okulova, Podgornykh, Paranchich, and Paranchich}}]{okulov:220}
\bibinfo{author}{\bibfnamefont{V.~I.} \bibnamefont{Okulov}},
  \bibinfo{author}{\bibfnamefont{A.~T.} \bibnamefont{Lonchakov}},
  \bibinfo{author}{\bibfnamefont{T.~E.} \bibnamefont{Govorkova}},
  \bibinfo{author}{\bibfnamefont{K.~A.} \bibnamefont{Okulova}},
  \bibinfo{author}{\bibfnamefont{S.~M.} \bibnamefont{Podgornykh}},
  \bibinfo{author}{\bibfnamefont{L.~D.} \bibnamefont{Paranchich}},
  \bibnamefont{and} \bibinfo{author}{\bibfnamefont{S.~Y.}
  \bibnamefont{Paranchich}}, \bibinfo{journal}{Low Temperature Physics}
  \textbf{\bibinfo{volume}{37}}, \bibinfo{pages}{220} (\bibinfo{year}{2011}).

\bibitem[{\citenamefont{Aleshkin et~al.}(2008)\citenamefont{Aleshkin,
  Gavrilenko, Odnoblyudov, and
  Yassievich}}]{springerlink:10.1134/S1063782608080034}
\bibinfo{author}{\bibfnamefont{V.}~\bibnamefont{Aleshkin}},
  \bibinfo{author}{\bibfnamefont{L.}~\bibnamefont{Gavrilenko}},
  \bibinfo{author}{\bibfnamefont{M.}~\bibnamefont{Odnoblyudov}},
  \bibnamefont{and}
  \bibinfo{author}{\bibfnamefont{I.}~\bibnamefont{Yassievich}},
  \bibinfo{journal}{Semiconductors} \textbf{\bibinfo{volume}{42}},
  \bibinfo{pages}{880} (\bibinfo{year}{2008}).

\bibitem[{\citenamefont{Dorokhin et~al.}(2010)\citenamefont{Dorokhin, Zaitsev,
  Brichkin, Vikhrova, Danilov, Zvonkov, Kulakovskii, Prokof'eva, and
  Sholina}}]{springerlink:10.1134/S1063783410110144}
\bibinfo{author}{\bibfnamefont{M.}~\bibnamefont{Dorokhin}},
  \bibinfo{author}{\bibfnamefont{S.}~\bibnamefont{Zaitsev}},
  \bibinfo{author}{\bibfnamefont{A.}~\bibnamefont{Brichkin}},
  \bibinfo{author}{\bibfnamefont{O.}~\bibnamefont{Vikhrova}},
  \bibinfo{author}{\bibfnamefont{Y.}~\bibnamefont{Danilov}},
  \bibinfo{author}{\bibfnamefont{B.}~\bibnamefont{Zvonkov}},
  \bibinfo{author}{\bibfnamefont{V.}~\bibnamefont{Kulakovskii}},
  \bibinfo{author}{\bibfnamefont{M.}~\bibnamefont{Prokof'eva}},
  \bibnamefont{and} \bibinfo{author}{\bibfnamefont{A.}~\bibnamefont{Sholina}},
  \bibinfo{journal}{Physics of the Solid State} \textbf{\bibinfo{volume}{52}},
  \bibinfo{pages}{2291} (\bibinfo{year}{2010}).

\bibitem[{\citenamefont{Zaitsev et~al.}(2010)\citenamefont{Zaitsev, Dorokhin,
  Brichkin, Vikhrova, Danilov, Zvonkov, and
  Kulakovskii}}]{springerlink:10.1134/S0021364009220056}
\bibinfo{author}{\bibfnamefont{S.}~\bibnamefont{Zaitsev}},
  \bibinfo{author}{\bibfnamefont{M.}~\bibnamefont{Dorokhin}},
  \bibinfo{author}{\bibfnamefont{A.}~\bibnamefont{Brichkin}},
  \bibinfo{author}{\bibfnamefont{O.}~\bibnamefont{Vikhrova}},
  \bibinfo{author}{\bibfnamefont{Y.}~\bibnamefont{Danilov}},
  \bibinfo{author}{\bibfnamefont{B.}~\bibnamefont{Zvonkov}}, \bibnamefont{and}
  \bibinfo{author}{\bibfnamefont{V.}~\bibnamefont{Kulakovskii}},
  \bibinfo{journal}{JETP Letters} \textbf{\bibinfo{volume}{90}},
  \bibinfo{pages}{658} (\bibinfo{year}{2010}).

\bibitem[{\citenamefont{Lucovsky}(1965)}]{Lucovsky}
\bibinfo{author}{\bibfnamefont{G.}~\bibnamefont{Lucovsky}},
  \bibinfo{journal}{Sol.St.Comm} \textbf{\bibinfo{volume}{3}},
  \bibinfo{pages}{299} (\bibinfo{year}{1965}).

\bibitem[{\citenamefont{Bardeen}(1961)}]{PhysRevLett.6.57}
\bibinfo{author}{\bibfnamefont{J.}~\bibnamefont{Bardeen}},
  \bibinfo{journal}{Phys. Rev. Lett.} \textbf{\bibinfo{volume}{6}},
  \bibinfo{pages}{57} (\bibinfo{year}{1961}).

\bibitem[{\citenamefont{Rozhansky et~al.}(2012)\citenamefont{Rozhansky,
  Averkiev, and L\"ahderanta}}]{PhysRevB.85.075315}
\bibinfo{author}{\bibfnamefont{I.~V.} \bibnamefont{Rozhansky}},
  \bibinfo{author}{\bibfnamefont{N.~S.} \bibnamefont{Averkiev}},
  \bibnamefont{and}
  \bibinfo{author}{\bibfnamefont{E.}~\bibnamefont{L\"ahderanta}},
  \bibinfo{journal}{Phys. Rev. B} \textbf{\bibinfo{volume}{85}},
  \bibinfo{pages}{075315} (\bibinfo{year}{2012}).

\bibitem[{\citenamefont{Schneider et~al.}(1987)\citenamefont{Schneider,
  Kaufmann, Wilkening, Baeumler, and K\"ohl}}]{PhysRevLett.59.240}
\bibinfo{author}{\bibfnamefont{J.}~\bibnamefont{Schneider}},
  \bibinfo{author}{\bibfnamefont{U.}~\bibnamefont{Kaufmann}},
  \bibinfo{author}{\bibfnamefont{W.}~\bibnamefont{Wilkening}},
  \bibinfo{author}{\bibfnamefont{M.}~\bibnamefont{Baeumler}}, \bibnamefont{and}
  \bibinfo{author}{\bibfnamefont{F.}~\bibnamefont{K\"ohl}},
  \bibinfo{journal}{Phys. Rev. Lett.} \textbf{\bibinfo{volume}{59}},
  \bibinfo{pages}{240} (\bibinfo{year}{1987}).

\bibitem[{\citenamefont{Zaitsev}(2012)}]{zaitsev:399}
\bibinfo{author}{\bibfnamefont{S.~V.} \bibnamefont{Zaitsev}},
  \bibinfo{journal}{Low Temperature Physics} \textbf{\bibinfo{volume}{38}},
  \bibinfo{pages}{399} (\bibinfo{year}{2012}).

\bibitem[{\citenamefont{Korenev et~al.}(2012)\citenamefont{Korenev, Akimov,
  Zaitsev, Sapega, Langer, Yakovlev, Danilov, and Bayer}}]{Korenev}
\bibinfo{author}{\bibfnamefont{V.~L.} \bibnamefont{Korenev}},
  \bibinfo{author}{\bibfnamefont{I.~A.} \bibnamefont{Akimov}},
  \bibinfo{author}{\bibfnamefont{S.~V.} \bibnamefont{Zaitsev}},
  \bibinfo{author}{\bibfnamefont{V.~F.} \bibnamefont{Sapega}},
  \bibinfo{author}{\bibfnamefont{L.}~\bibnamefont{Langer}},
  \bibinfo{author}{\bibfnamefont{D.~R.} \bibnamefont{Yakovlev}},
  \bibinfo{author}{\bibfnamefont{Y.~A.} \bibnamefont{Danilov}},
  \bibnamefont{and} \bibinfo{author}{\bibfnamefont{M.}~\bibnamefont{Bayer}},
  \bibinfo{journal}{Nat. Commun.} \textbf{\bibinfo{volume}{3}},
  \bibinfo{pages}{959} (\bibinfo{year}{2012}).

\end{thebibliography}

\clearpage

\begin{figure}
  \leavevmode
 \centering\includegraphics[width=0.5\textwidth]{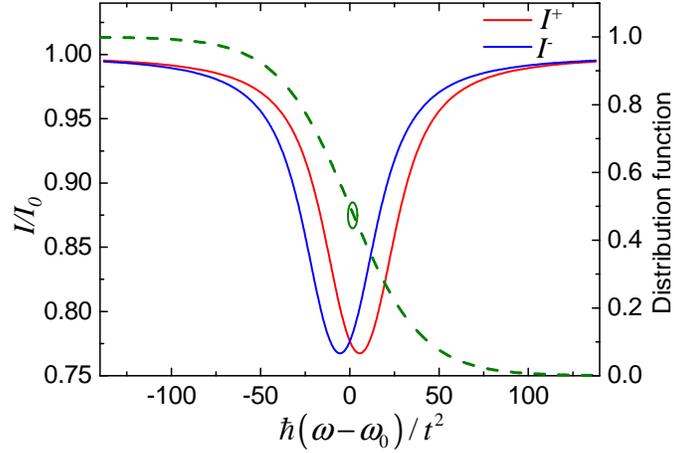}
 \caption{(Color online) Modification of the luminescence spectrum
 by tunneling configuration interaction. The integral polarization occurs when the carriers distribution function (dashed line) strongly varies
 in the vicinity of the configuration resonances. $\omega_0$ is the position of the resonance without bound level splitting.}
 \label{fig_Enhancement}
\end{figure}

\begin{figure}
  \leavevmode
 \centering\includegraphics[width=0.5\textwidth]{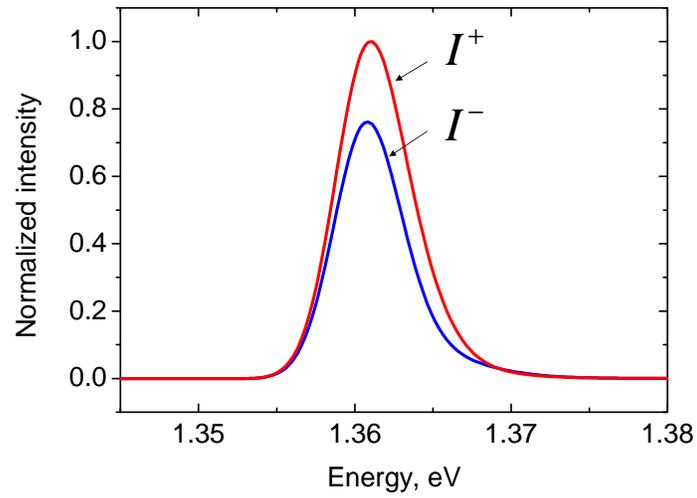}
 \caption{(Color online) An example of calculated luminescence spectra for the two circular polarizations.
The case of antiferromagnetic coupling implies $I^-$ corresponds to
($\sigma^+$) polarization while $I^+$ to ($\sigma^-$) polarization.
  The parameters used in calculations are given in the text. }
 \label{figSpectra}
\end{figure}



\end{document}